# Boron Clusters for Metal-Free Water Splitting

Masaya Fujioka[1,2] *, Haruhiko Morito[3], Melbert Jeem[4], Jeevan Kumar Padarti[5], Kazuki Morita[6], Taizo Shibuya[7], Masashi Tanaka[8], Yoshihiko Ihara[9], Shigeto Hirai[5] **

[1] Multi-Materials Research Institute, National Institute of Advanced Industrial Science and Technology (AIST), Nagoya 463-8560, Japan
[2] Research Institute for Electronic Science, Hokkaido University, Sapporo 001-0020, Japan
[3] Institute for Materials Research, Tohoku University, Sendai 980–8577, Japan
[4] Faculty of Engineering, Hokkaido University, Sapporo, 060-8628, Japan
[5] Faculty of Engineering, Kitami Institute of Technology, Kitami 090–8507, Japan
[6] Helmholtz-Zentrum Berlin für Materialien und Energie, 12489 Berlin, Germany
[7] System Platform Research Laboratories, NEC Corporation, Kawasaki, Kanagawa 211-8666, Japan
[8] Graduate School of Engineering, Kyushu Institute of Technology, Kitakyushu 804-8550, Japan
[9] Department of Physics, Faculty of Science, Hokkaido University, Sapporo 060-0810, Japan

Corresponding authors
* m.fujioka@aist.go.jp
** hirai@mail.kitami-it.ac.jp

## Abstract

Electron-deficient boron clusters are identified as a fundamentally new class of oxygen evolution reaction (OER) catalysts, entirely free of transition metals. Selective sodium extraction from $NaAlB_{14}$ and $Na_2B_{29}$ via high-pressure diffusion control introduces hole doping into $B_{12}$ icosahedral frameworks, resulting in OER activity exceeding that of $Co_3O_4$ by more than an order of magnitude, and exceptional durability under alkaline conditions. $B_{12}$ clusters are known for their superchaotropic character, which destabilizes hydrogen bonding in water. In this system, $H_2O$, instead of $OH^-$, preferentially adsorbs on the catalyst surface, suggesting a distinct OER pathway mediated by molecular water. This adsorption behavior contrasts with conventional transition-metal oxides and reflects the unique interfacial properties of the boron clusters. Density functional theory reveals unoccupied $p$ orbitals and unique local electric fields at the cluster surface, both of which could promote the water activation. These findings suggest a paradigm shift in OER catalysis, in which the unique interaction between $B_{12}$ clusters and water drives the reaction, replacing the conventional role of redox-active metals. Hole-doped boron clusters thus offer a promising platform for designing high-performance and durable water-splitting catalysts, opening new avenues for OER design beyond conventional transition-metal chemistry.

## 1. Introduction

The oxygen evolution reaction (OER) is a crucial reaction for emerging energy conversion technologies such as electrochemical water splitting [1, 2] and rechargeable metal-air batteries [3, 4]. In particular, the growing interest in water electrolyzers for hydrogen production, such as using the surplus electric power from renewable energy [5, 6], has led to extensive research for the enhancement of OER activity and durability. Moreover, recent discoveries of transition metal oxide-based catalysts [7, 8, 9, 10, 11, 12, 13, 14, 15, 16] in both alkaline and acidic solutions have greatly advanced our understanding of the OER, revealing new OER mechanisms [17, 18, 19], surface modification through new strategies [20, 21], construction of durable surface structure through strong electron-electron correlation [15, 22], and numerous OER descriptors [23, 24, 25].

Despite these advances, OER remains a key bottleneck in water electrolyzers and rechargeable metal-air batteries due to its sluggish kinetics arising from a multistep electron transfer process [4]. In addition, the high overpotential, together with the dynamic nature of the catalytic surfaces, leads to severe elemental dissolution and amorphization of the surface structure. Thus, finding OER catalysts with both high activity and stability is a crucial and challenging task.

Although the possible OER mechanisms are still under discussion, the necessity of redox-active transition metal centers as active sites has been widely accepted in this field, typically based on mechanisms such as the adsorbate evolution mechanism and the lattice oxygen-mediated mechanism [26, 27, 28, 29, 30]. At the same time, breakthroughs are needed for further advancement and a better understanding of the OER catalysis. In this aspect, metal-free systems or those dominated by non-oxides are of significant interest as OER catalysts [31, 32, 33, 34], yet they have remained largely unexplored because of the scarcity of accessible candidates, and the absence of transition-metal centers. Thus, the discovery of a highly efficient metal-free OER catalyst represents a major breakthrough, fundamentally challenging the conventional views of catalyst design.

In such circumstances, we have found an exceptional OER activity enhancement in boron cluster compounds, and therefore propose them as a new class of transition-metal-free OER catalysts. Boron, being electron-deficient, favors multi-center bonding schemes such as three-center two-electron (3c–2e) bonds, which give rise to stable and rigid cluster frameworks [35]. Notably, icosahedral $B_{12}$ clusters, a representative example, exhibit "superchaotropic behavior": being highly soluble in water while simultaneously disrupting the hydrogen-bonding network of the surrounding solvent [36]. This paradoxical combination of hydrophilic solubility and hydrophobic-like interfacial character imparts these clusters with unique and intriguing physicochemical properties.

On the other hand, boron-rich compounds exhibit a self-compensation effect[37, 38], which poses a major challenge for tuning their electronic structure. Unlike ion intercalation or heterovalent elemental substitution, this effect hinders effective modulation of charge-carrier concentration, as carrier introduction is often counteracted by the spontaneous formation of interstitial boron atoms or boron vacancies. This intrinsic self-regulating behavior tends to neutralize changes in carrier density, thereby limiting the rational design of electronic functionality in borides.

Meanwhile, we have advanced a materials design strategy based on diffusion control, which enables the modulation of chemical compositions that are unattainable under thermodynamic equilibrium. Our synthesis techniques, including PDII [39, 40], ADC[41], and HPDC[42, 43], facilitate compositional tuning through the extraction, introduction, and substitution of elements while retaining the fundamental framework of the host structure[44]. In this study, the HPDC method was employed to selectively extract Na from compounds such as $NaAlB_{14}$[45, 46] and $Na_2B_{29}$[47], thereby enabling effective hole doping within the boron cluster framework formed by icosahedral $B_{12}$ units[42]. Such non-equilibrium synthesis strategies allow the formation of metastable structures inaccessible under thermodynamic equilibrium, potentially leading to novel catalytic and electronic functionalities. By merging this advanced synthesis approach with the distinctive bonding state (3c–2e) and anomalous physicochemical properties (such as superchaotropic properties) of $B_{12}$ clusters, this work opens up a new paradigm in the design of transition-metal-free OER catalysts.

## 2 Results and discussion
### 2.1. OER properties

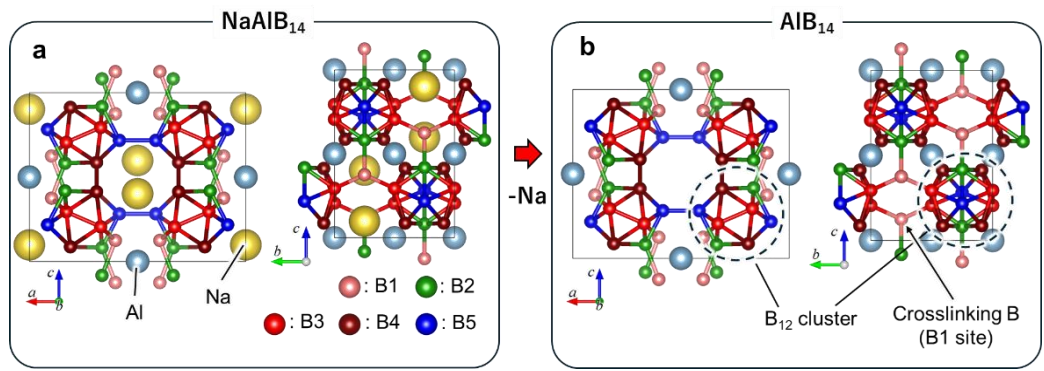

Figure 1 | Crystal structures of $NaAlB_{14}$ and $AlB_{14}$.
(a) Crystal structure of $NaAlB_{14}$ showing $B_{12}$ icosahedra. (b) Crystal structure of $AlB_{14}$ after Na extraction.

Figure 1a shows the crystal structure of NaAlB$_{14}$, which features B$_{12}$ icosahedral clusters formed by B atoms at the B2–B5 sites. These structures were visualized using Visualization for Electronic and STructural Analysis (VESTA)[48]. These clusters are interconnected via B atoms at the B1 sites, resulting in a three-dimensional covalent network. Na and Al atoms are accommodated within this boron framework. Figure 1b depicts the structure of AlB$_{14}$, obtained by selective Na extraction from NaAlB$_{14}$. The synthesis procedure has been described in detail in our previous studies[49].

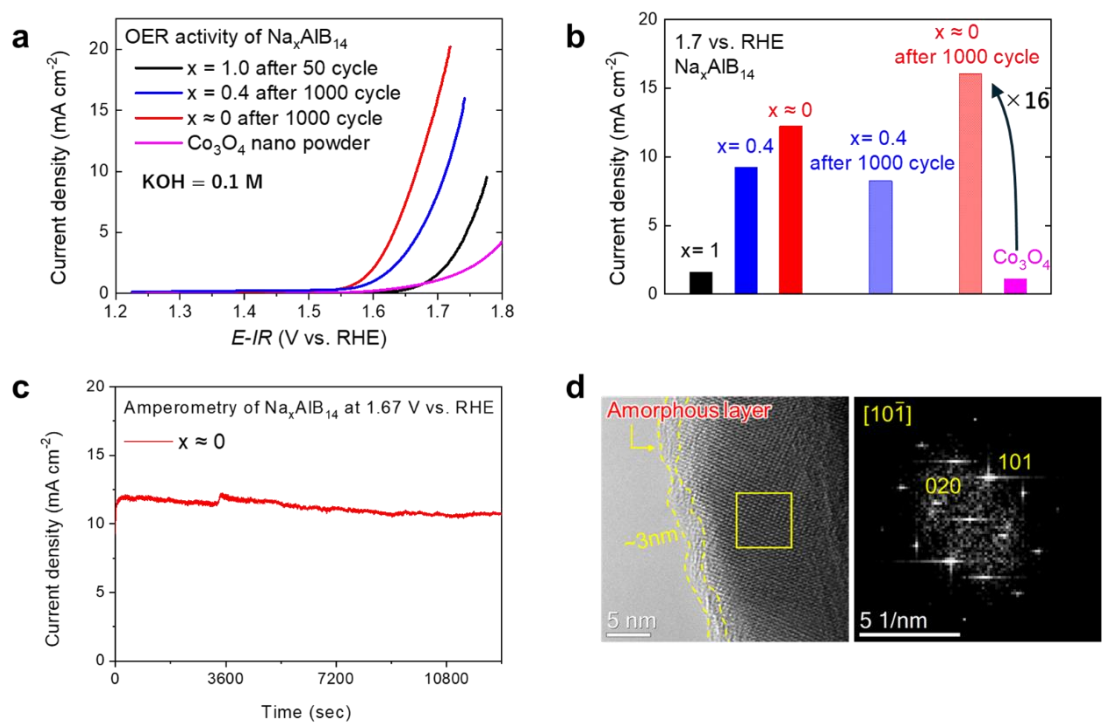

**Figure 2 | OER performance and structural characterization of Na$_x$AlB$_{14}$.**
(a) Linear sweep voltammograms of Na$_x$AlB$_{14}$ (x = 0, 0.4, 1) and Co$_3$O$_4$ nanoparticles in 0.1 M KOH aqueous solution at the scan rate of 10 mV/s. Note that for x = 1, measurements were limited to 50 cycles due to electrode delamination. (b) OER current density at 1.7 V vs. RHE for each catalyst. (c) Chronoamperometric curve of Na$_x$AlB$_{14}$ (x ≈ 0) at 1.67 V vs. RHE. (d) High-resolution TEM (HRTEM) image of an AlB$_{14}$ particle, marking the approximately 3 nm amorphous layer. Right image shows the Fast Fourier Transform (FFT) pattern obtained from the yellow-marked crystalline region.

Figures 2a and 2b summarize the OER performance of Na$_x$AlB$_{14}$ (x = 0, 0.4, 1) measured in 0.1 M KOH aqueous solution. A clear enhancement in OER activity is observed with decreasing Na content. In particular, AlB$_{14}$ with nearly complete Na removal (x ≈ 0) exhibits exceptionally high activity, up to 16 times higher than that of commercial $Co_3O_4$ nanopowder. Notably, both the x = 0.4 and x ≈ 0 samples show further improvement in OER performance after 1000 electrochemical cycles, as shown in Figures S1a and S1b. Although high current density tends to come from high surface area, which is generally achieved by nanoscale particle size, the $Na_xAlB_{14}$ samples in this study were used in their as-ground form without further size refinement, resulting in a broad particle size distribution, including coarse grains exceeding several tens of micrometers (Figure S1c). Remarkably, despite their relatively large particle size, these materials exhibit outstanding OER activity. Similar behavior has only rarely been observed in a few transition-metal-based systems[15, 20, 50]. The observation of such high activity in a transition-metal-free boron cluster compound is particularly striking, demonstrating that this system provides a promising playground for the development of OER catalysis. Furthermore, future improvements in OER performance may be achieved by deliberate strategies such as nanoscale refinement of the boron cluster catalyst or formation of composites with other OER-facourable nanoparticles to increase the electrochemically active surface area and active sites.

Figure 2c shows the chronoamperometric curve of $Na_xAlB_{14}$ (x ≈ 0), representing the long-term durability of the catalyst under OER conditions. At an applied potential of 1.67 V vs. RHE, a current density of 10 mA/cm$^2$ is well-maintained over an extended period. On the contrary, for transition metal oxide catalysts, the harsh oxidative conditions of the OER often lead to surface amorphization, driven by the leaching of metal species from the surface. This process gives rise to chemical and structural disparities between the catalytic surface and the underlying bulk layer, and such leaching is ultimately regarded as a key factor contributing to the deterioration of the catalyst's long-term durability. It remains one of the central challenges in the development of stable OER electrocatalysts[51].

The HRTEM image in Figure 2d presents the surface of $AlB_{14}$ after 1000 OER cycles, revealing approximately 3 nm thick amorphous layer. Meanwhile, the FFT pattern shows [00$\bar{1}$] view direction of the $AlB_{14}$ crystal, in which $d_{101}$ = 4.53 Å and $d_{020}$ = 5.32 Å signify no peculiar change in the $AlB_{14}$ lattice parameters[42] (Figure S2a and S2b).

Notably, STEM-EDS analysis (Figure S3) shows that the chemical composition of the amorphous layer is nearly identical to that of the crystalline layer, which is predominantly composed of boron. Additional compositional EDS data from the overview map are provided in Figure S4 and S5. No evidence of the elemental dissolution or surface layer oxidation was observed during the OER.

Unlike $BaIr_{1-x}Mn_xO_3$, where minimal changes of surface composition are observed before and after OER[20], boron cluster compounds exhibit a fundamentally distinct mechanism underlying their surface stability. The crystal structure of $NaAlB_{14}$ is composed of a three-dimensional network of covalently bonded boron atoms. This robust boron framework has been reported to remain structurally intact even upon the extraction of Na and Al, indicating a high degree of resistance to ion leaching[42]. This feature is not typically observed in oxide-based catalysts, whose frameworks are generally formed through more ionic bonding and are thus more susceptible to structural disruption upon cation leaching.

Furthermore, as shown in Figures S6a and S6c, this amorphous surface layer is already present prior to the surface treatment during the OER and remains largely unchanged even after the OER. For this reason, we deduce that this is an intrinsic structural feature resulting from extensive Na extraction achieved through diffusion control by HPDC. The HRTEM images (Figure S6) confirm that there was no lattice parameter change after the OER. In contrast, the sample with $x = 0.4$ exhibits a crystalline surface before the OER, but after the OER, Na-rich precipitates emerge, and the surface amorphization starts to proceed. This precipitation is possibly due to the Na dissolution that leads to the partial surface reconstruction.

Thus, the covalently bonded boron framework is inherently resilient to cation leaching, owing to its robust network structure. Notably, the amorphous region observed at the surface is far from being fully disordered; the local geometry of the $B_{12}$ icosahedral clusters is likely preserved, as suggested by the well-established structural robustness and exceptional stability of $B_{12}$ clusters[52, 53]. Indeed, prior studies employing neutron PDF analysis have demonstrated that even in amorphous boron, the local $B_{12}$ geometry remains intact despite the loss of long-range order[54].

Consequently, the framework formed by boron clusters and their bridging boron atoms offers a structural advantage in accommodating cation extraction through three-dimensional interconnected channels. Even when an amorphous layer forms at the surface, the $B_{12}$ cluster units are expected to remain intact, thereby preserving their unique physicochemical properties.

## 2.2. Sodium Extraction and Defect Reconfiguration

The enhancement of OER activity upon Na extraction provides key insights into the catalytic mechanism of this system. Boron-rich borides are known to accommodate a large number of structural defects and interstitial boron atoms, raising the possibility that such defective sites may serve as active centers. To evaluate this possibility, electron spin resonance (ESR) measurements were conducted to quantify the total number of unpaired electrons at different Na concentrations. As shown in Figures 3a and 3b, the ESR signal becomes sharper and the number of unpaired electrons decreases as the Na content is reduced. This inverse correlation between unpaired electron density and OER activity indicates that boron defects are unlikely to function as direct active sites.

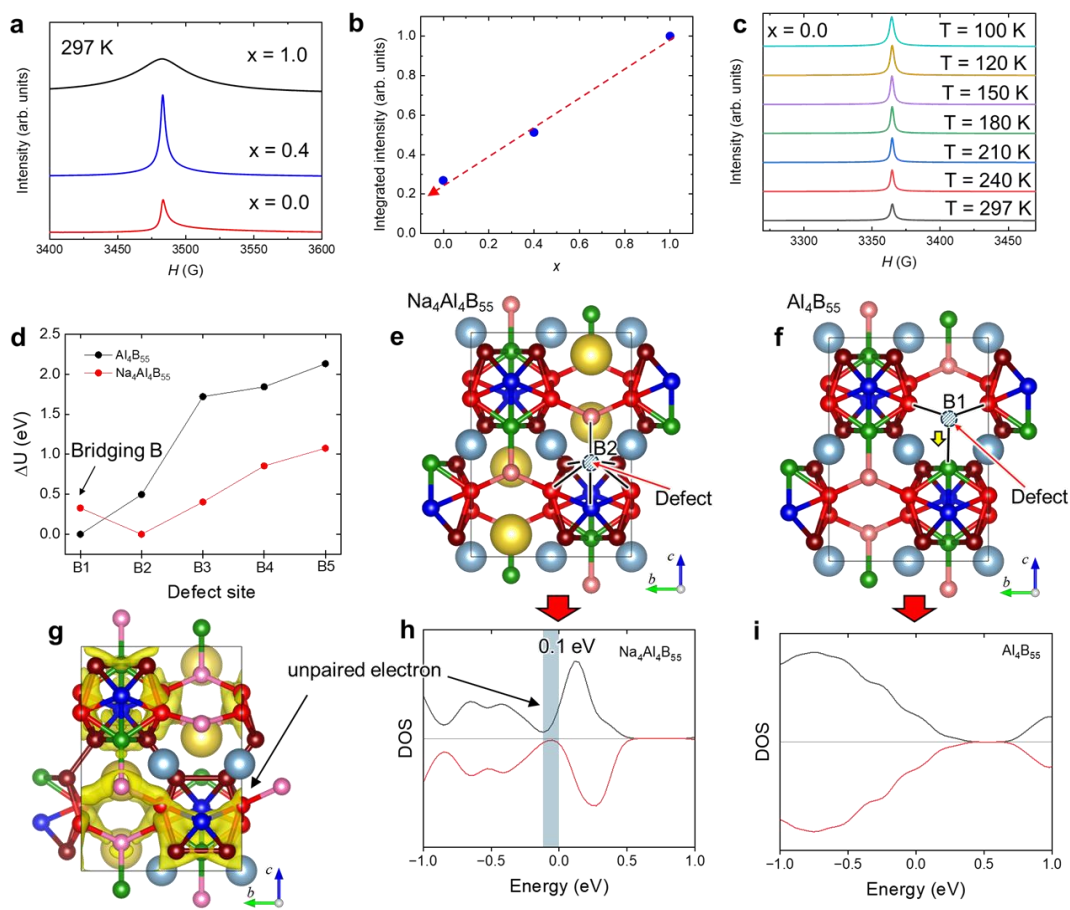

**Figure 3 | ESR analysis and defect evolution in $Na_xAlB_{14}$.**
(a) ESR spectra of $Na_xAlB_{14}$ (x = 0, 0.4, 1). (b) Integrated ESR signal intensity as a function of Na content. (c) Temperature dependence of the ESR spectrum for $AlB_{14}$. (d) Relative stability of defect-introduced structures before and after Na removal. (e) Crystal structure of $Na_4Al_4B_{55}$ with a defect at the B2 site. (f) Crystal structure of $Al_4B_{55}$ with a defect at the B1 site. (g) Partial charge density of $Na_4Al_4B_{55}$, visualized for electronic states within 0.1 eV below the Fermi level. (h) Calculated density of states (DOS) near the Fermi level for $Na_4Al_4B_{55}$. Black and red lines denote the up and down spin respectively. (i) Calculated DOS for $Al_4B_{55}$.

The observed sharpening of the ESR peak upon Na removal can be attributed to two potential factors: (i) enhanced electronic conductivity, which may lead to exchange narrowing effects[55], and (ii) structural factors that result in the preferential retention of specific defect sites. As shown in Figure 3c, $AlB_{14}$ maintains a sharp ESR signal even at low temperatures, despite exhibiting a steep increase in resistivity (Figure S4), indicating that the spectral sharpening is primarily of structural origin.

To further probe the nature of defect formation, DFT calculations[56, 57] were performed to evaluate the relative total energies of structures with a single boron vacancy introduced at each of the B1–B5 sites within the covalent boron framework (Figure 3d). In $NaAlB_{14}$, the B2 site is the most energetically favorable for defect formation, whereas in the Na-deficient system, the lowest-energy site shifts to B1. This change suggests a structural rearrangement in which bridging B1 atoms are mobilized to compensate for vacancies in the $B_{12}$ clusters following Na removal (Figures 3e and 3f).

Such structural evolution is consistent with the intrinsic bonding nature of electron-deficient boron frameworks. As discussed above, boron atoms tend to cluster in response to their electron deficiency, forming multi-centered bonds such as the 3c–2e type. When metal content increases and additional electrons are introduced into the system, the driving force for cluster formation is weakened, resulting in a rise in defect concentration. Ultimately, as seen in compounds like $MgB_2$, boron adopts conventional 2c–2e bonding in planar $sp^2$ networks. Conversely, the hole doping induced by Na extraction in $AlB_{14}$ further stabilizes the $B_{12}$ cluster framework by reinforcing its electron-deficient character.

This structural rearrangement, in which B1-site boron atoms migrate to compensate for vacancies at B2 sites upon Na extraction, is supported by electronic structure calculations. Figures 3h and 3i show the calculated density of states (DOS) near the Fermi level for $Na_4Al_4B_{55}$ and $Al_4B_{55}$ models with boron vacancies at the B2 and B1 sites, respectively. In the Na-containing model with a vacancy at the B2 site, a clear spin polarization appears at the Fermi level, indicative of unpaired electrons. In contrast, in the Na-deficient model with the vacancy relocated to the B1 site, the spin polarization vanishes, reflecting the elimination of unpaired electronic states. The observed decrease in unpaired electrons upon defect migration is consistent with the ESR results shown in Figure 3a.

Figure 3g visualizes the partial charge density corresponding to the electronic states just below the Fermi level, as indicated in Figure 3h. The boron clusters, which share electrons through multi-center, multi-electron bonding, are known to promote delocalized electronic states across the extended framework[58]. The widespread distribution of unpaired electrons by the defect of the cluster observed in Figure 3g is therefore considered to originate from this characteristic bonding nature. This delocalization is also consistent with the broad ESR signal observed in the Na-rich sample (Figure 3a).

Furthermore, the loss of these bridging boron atoms introduces increased configurational disorder among the $B_{12}$ clusters, making it increasingly difficult to maintain structural periodicity near the surface. In $NaAlB_{14}$, surface amorphization proceeds with decreasing Na content, regardless of the surface treatment during OER (Figure S6a and S6b). This phenomenon is likely related to the loss of bridging boron atoms that accompanies Na extraction.

## 2.3. Surface Water Adsorption Dominance in $Na_xAlB_{14}$ under Alkaline Conditions

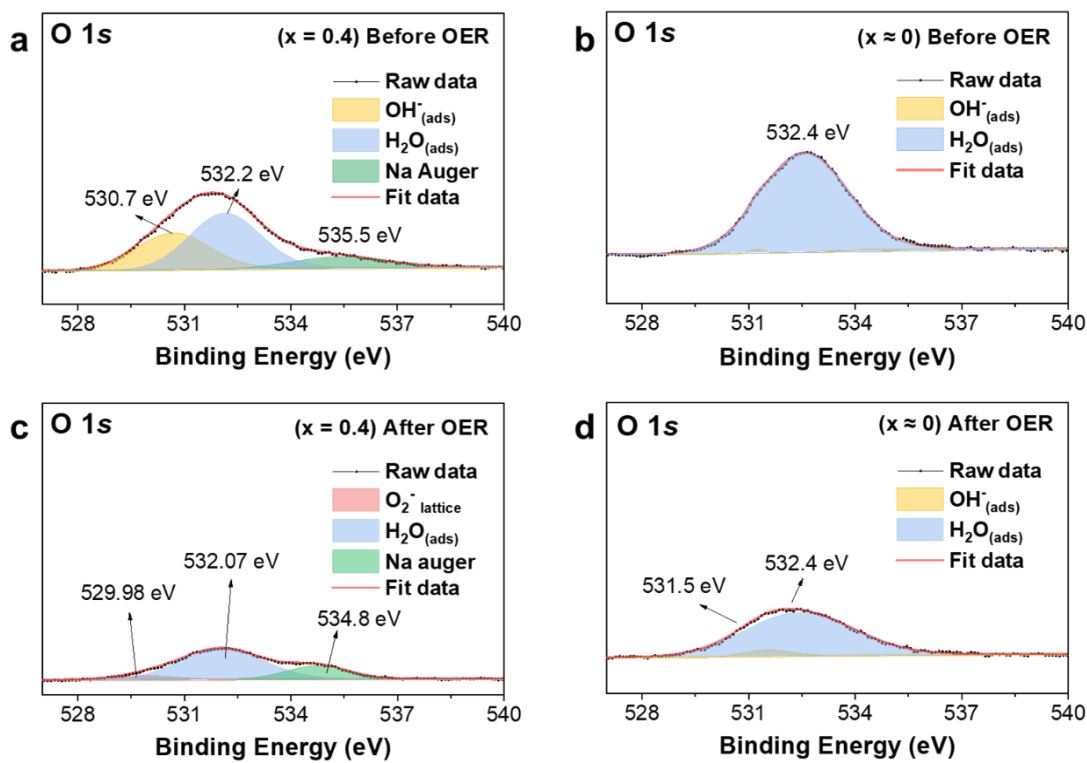

**Figure 4 | Surface adsorption behavior of $Na_xAlB_{14}$ under alkaline conditions**
XPS spectra of the x = 0.4 sample (a) before and (b) after OER, and of the x ≈ 0 sample (c) before and (d) after OER.

Under alkaline conditions, oxygen evolution reaction (OER) on conventional transition metal oxide catalysts typically begins with the adsorption of hydroxide ions ($OH^-$) onto metal active sites. Catalysts that offer surfaces favorable for $OH^-$ adsorption generally show higher OER activity. In contrast, Na$_x$AlB$_{14}$ exhibits high OER activity despite predominantly adsorbing H$_2$O instead of OH$^-$ on its surface. Figure 4a–4d present O 1s XPS spectra obtained from the same TEM samples shown in Figure S6.

As shown in Figures 4a and 4b, the x = 0.4 sample displays partial OH$^-$ adsorption before the surface treatment during the OER. After the reaction, however, OH$^-$ species are almost entirely removed, and H$_2$O becomes the dominant adsorbate. This suggests that the reaction proceeds via a mechanism initiated by water adsorption, indicating that the reaction may follow a distinct route from the conventional OH$^-$ adsorption-based mechanism.

In contrast, for Na-deficient AlB$_{14}$ (x ≈ 0), Figures 4c and 4d show that H$_2$O remains the dominant adsorbate both before and after the surface treatment during the OER. This suggests that the Na-deficient surface, where B$_{12}$ clusters are more stabilized, offers a highly favorable environment for water adsorption even under alkaline conditions (0.1 M KOH). This unconventional adsorption behavior is likely associated with the superchaotropic nature of the B$_{12}$ clusters and represents a marked departure from the typical OER pathways observed in transition metal oxide catalysts.

## 2.4. Hole doping as a general strategy for activating B$_{12}$ cluster-based borides

As shown in Figure 5a, the removal of Na from NaAlB$_{14}$, corresponding to hole doping, leads to a pronounced enhancement in OER activity. In contrast, electron doping via substitution of Na with Mg, as in MgAlB$_{14}$[59], significantly suppresses the catalytic performance. Since both compounds share an identical B$_{12}$ cluster framework, these results indicate that hole doping serves as the key trigger for enhancing OER activity.

Figure 5b extends this concept to another boron-rich compound, Na$_2$B$_{29}$, which also features a B$_{12}$-based cluster structure. The sample consists of a mixture of Na$_x$B$_{29}$ regions with varying sodium content: some where most of the Na has been extracted, and others where it remains largely intact; the average Na content is approximately x = 1.2 (see Figure S7 for details). The enhancement in OER activity observed upon partial Na removal follows the same trend as NaAlB$_{14}$.

These results demonstrate that the catalytic enhancement driven by hole doping is not confined to NaAlB$_{14}$ but is likewise realized in aluminum-free Na$_2$B$_{29}$. This supports the central role of the B$_{12}$ cluster, instead of the surrounding cations, as the key structural unit responsible for catalytic activity.

Taken together, these findings establish hole doping in B$_{12}$-based borides as a versatile strategy for the development of high-performance OER catalysts. While the precise mechanistic underpinnings require further elucidation, the data presented here point to increasing electron deficiency within $B_{12}$ clusters as a critical driver of catalytic function. This insight positions boron-rich borides as promising candidates for sustainable water-splitting technologies and future hydrogen production.

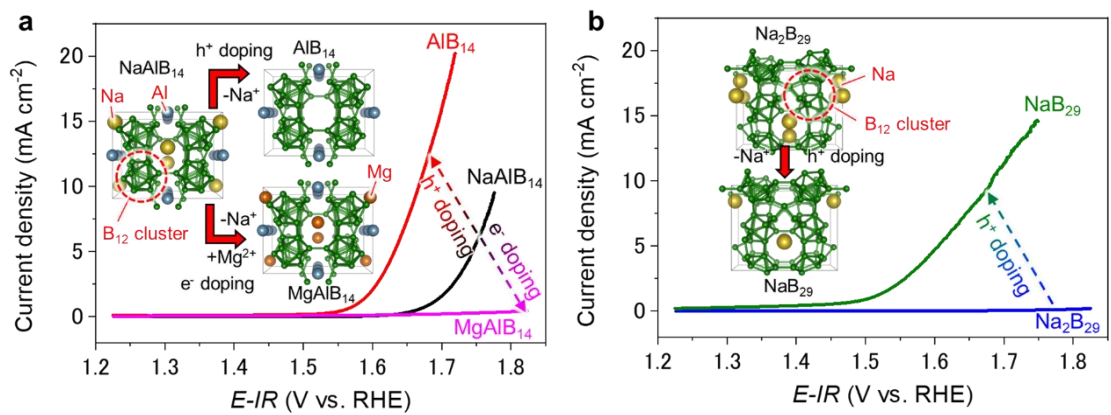

**Figure 5 | Role of hole doping in $B_{12}$-based OER catalysts**
(a) OER activity of $AlB_{14}$, $NaAlB_{14}$ and $MgAlB_{14}$. (b) OER activity of $Na_2B_{29}$, Na extracted $Na_xB_{29}$.

## 2.5. Possible Adsorption Modes on $B_{12}$ Clusters via DFT

To investigate how water interacts with the surfaces of $B_{12}$ cluster-based materials, symmetric slab models of $AlB_{14}$ and $NaAlB_{14}$ exposing the (001) facet were constructed based on 1×1×2 supercells with a 15 Å vacuum layer. Since the outermost surface of $Na_xAlB_{14}$ becomes amorphous upon Na extraction, precise modeling of adsorption sites is inherently challenging. However, previous studies demonstrated that $B_{12}$ clusters retain their icosahedral geometry even in amorphous boron[54]. This indicates that the electronic structure near the Fermi level is likely to preserve features of the crystalline phase, provided that the local bonding environment around the $B_{12}$ clusters remains largely intact.

Based on this assumption, Figures 6a and 6b compare the Bader charges of surface-exposed $B_{12}$ clusters and bridging boron atoms with those in the bulk. At the surface, specific boron sites (e.g., B3, B4, and B5) within the $B_{12}$ clusters acquire a partially positive charge, a feature that becomes more pronounced upon Na extraction. This local arrangement of adjacent positively and negatively charged sites may give rise to an inhomogeneous local electric field, which serves as an electrostatic driving force for attracting water molecules.

Figures 6c and 6d show the calculated DOS for bulk and surface models. In Na-deficient $AlB_{14}$, the Fermi level lies within the valence band in both cases, indicating a metallic or degenerate semiconducting character. In contrast, NaAlB$_{14}$ exhibits a band gap in both models; however, the gap from the surface model is considerably narrower than that from the bulk model. Partial charge density maps of unoccupied states within 1 eV above the Fermi level, shown in Figures 6e and 6f, reveal unoccupied $p$ orbitals extending outward from the surface. These orbitals likely act as electron-accepting sites for adsorbates, including lone pairs from water or hydroxide. The absence of a band gap in AlB$_{14}$ suggests that such electron transfer processes may occur more readily than in NaAlB$_{14}$, potentially enhancing adsorption and facilitating OER-related reactions.

While the precise identity of active adsorption sites remains unclear due to the disordered surface structure, the DFT results point to two distinct modes of interaction between water molecules and B$_{12}$ clusters: (i) physical adsorption driven by electrostatic attraction, and (ii) chemical adsorption via Lewis acid–base interactions involving unoccupied $p$ orbitals. These features may be associated with the distinct superchaotropic character of B$_{12}$ clusters. Although many mechanistic aspects remain to be clarified, the present findings provide important clues that are expected to guide future investigations toward a deeper understanding of catalytic behavior in boron-rich systems.

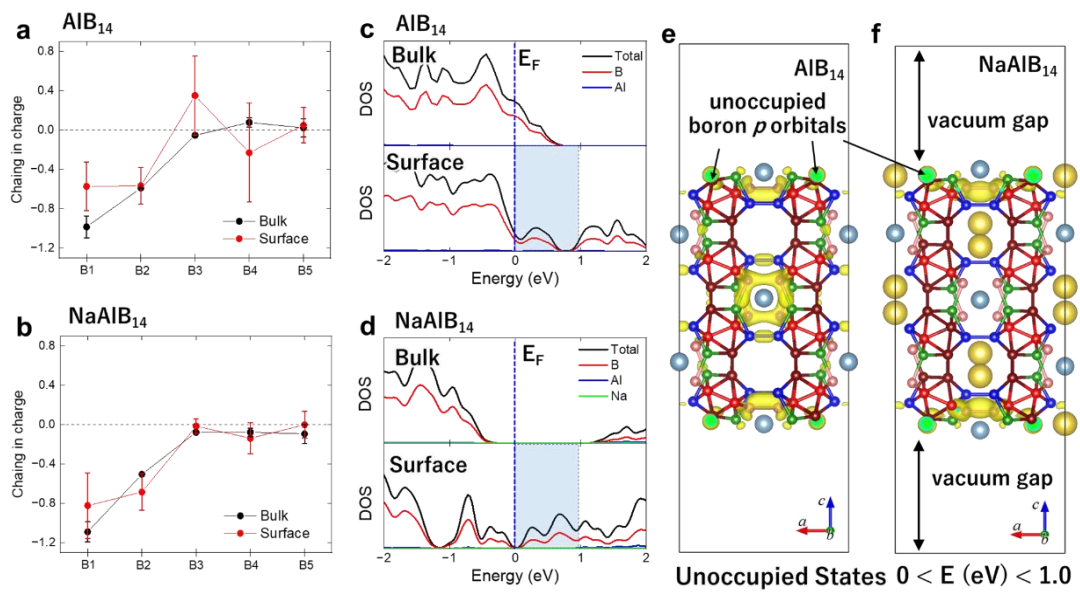

**Figure 6 | DFT insights into the origin of OER activity**

Change in the charge of each B site in (a) AlB$_{14}$ and (b) NaAlB$_{14}$, estimated using Bader charge analysis. Density of states (DOS) near the Fermi level for the bulk and surface regions of (c) AlB$_{14}$ and (d) NaAlB$_{14}$. Partial charge density distribution above the Fermi level (within 1 eV) for the surface of (e) AlB$_{14}$ and (f) NaAlB$_{14}$. The isosurface threshold for visualizing the spatial distribution of the $p$ orbitals is set to 0.01 e/Å$^3$.

## 3. Conclusion

This work establishes electron-deficient $B_{12}$ icosahedral clusters as a previously unrecognized and transition-metal-free platform for oxygen evolution reaction (OER) catalysis. We demonstrate that hole doping, induced by sodium deintercalation from $NaAlB_{14}$ and $Na_2B_{29}$, substantially enhances OER activity, whereas electron doping via magnesium substitution has the opposite effect. These findings identify boron clusters, instead of the transition metals, as the key catalytic unit governing OER behavior. In particular, nearly Na-free $AlB_{14}$ exhibits outstanding catalytic performance in 0.1 M KOH, showing up to 16 times higher OER activity than commercial $Co_3O_4$ nanoparticles. A current density of 10 mA/cm$^2$ is sustained over extended periods at 1.7 V vs. RHE, highlighting the long-term stability of the catalyst under alkaline conditions.

Although surface amorphization is commonly observed during OER and often leads to catalyst degradation, the $B_{12}$-based compounds studied here exhibit excellent durability. This resilience arises from the structural nature of the boron cluster compounds, which permits cation leaching while preserving the integrity of the covalent structural network. Notably, an amorphous layer was also observed on the surface of Na-deficient $AlB_{14}$, but this transformation does not result from OER-induced degradation. Instead, it is an intrinsic surface reconstruction triggered by Na extraction, reflecting a material-specific response of the boron framework.

These DFT-derived insights, together with the uniquely superchaotropic nature of the $B_{12}$ clusters, are closely related to the observed unusual preference for $H_2O$ adsorption in alkaline media. This behavior likely reflects a distinct mechanistic pathway, fundamentally different from that of conventional transition-metal-based OER catalysts. It should be emphasized that this is a newly explored catalytic system, and a comprehensive mechanistic understanding will require further investigation. Nonetheless, the present study establishes key structure–function relationships and highlights the central role of electron-deficient $B_{12}$ clusters in promoting OER. These insights offer a conceptual departure from traditional transition metal-based catalysts and pave the way for future research into boron-rich compounds as a new platform for water-splitting catalysis.

## Method

### Sample preparation and characterization

$NaAlB_{14}$ and $Na_2B_{29}$ powders were synthesized by sintering under Na vapor, as described in a previous report[45]. To ensure strong intergranular connectivity suitable for high-pressure diffusion control (HPDC) treatment[60], the as-synthesized powders were annealed at 900 °C for 2 hours under a pressure of 4 GPa. The Na$_x$AlB$_{14}$ samples with varying sodium content (x = 0, 0.4, and 1) were prepared and characterized as reported previously[49].

A powder sample of Na$_2$B$_{29}$ was prepared by heating a crystalline B powder in Na vapor (vapor-solid reaction.) The metal Na (Nippon Soda Co., Ltd., purity; 99.95%) and crystalline B powder (Hirano Seizaemon Shouten Co., Ltd., purity; 99.4%, particle size; <0.5 mm) were weighed at a molar ratio of Na: B = 1: 1, and placed in a BN crucible (Zikusu Industry Co., Ltd., 99.7%, 8.5 mm outer diameter, 6.5 mm inner diameter, 18 mm depth). The crucible was sealed in a stainless steel container in Ar atmosphere, and heated at 1100 °C for 12 h. When heating the container above 1000 °C, both ends of the container shown in the reference[45] were sealed with caps to prevent Na leakage. After cooling, the crucible was taken out from the container, and Na remaining in the crucible were reacted and removed in the order of 2-propanol and ethanol. Furthermore, the sample was washed with pure water to remove water-soluble compounds such as alkoxide produced by the reaction of Na and alcohol.

Sodium extraction from Na$_2$B$_{29}$ was carried out using the HPDC technique at 550 °C for 20 hours under 1 GPa. The sample stack in the HPDC cell consisted of carbon, Na$_2$B$_{29}$, Y-type zeolite, and carbon-loaded Y-type zeolite layers, as illustrated in Figure S4a. The current flow through the stack was continuously monitored under an applied voltage of 5 V to assess the progression of the Na-extraction process (Figure S4b). MgAlB$_{14}$ powder was also prepared by heating Mg (Kojundo Chemical Laboratory, 99.9%), Al powder (Kojundo Chemical Laboratory, 99.99%), and crystalline B powder (molar ratio Mg: Al: B = 1: 1: 14.2) at 1000 °C for 24 h with same container.

Powder X-ray diffraction (XRD) measurements were conducted using MiniFlex 600 with D/teX Ultra (Rigaku), and the sodium content was evaluated via SEM-EDS using JCM-6000 with EX-37001 (JEOL). The X-band ESR spectra were measured using EMXplus (Bruker). Core-level spectra from X-ray photoemission spectroscopy (XPS) were obtained by a PHI 5000 VersaProbe spectrometer using Al-Ka radiation at room temperature. For all the core-level spectra, Shirley background was subtracted from the raw data for the XPS analysis. For scanning/transmission electron microscopy (S/TEM) analyses, samples were deposited onto a perforated amorphous carbon film supported Mo grid (EM Japan, Co., Ltd). High resolution S/TEM and STEM-EDS were carried out using JEM ARM-200F (JEOL). The microscope was operated at 200 kV, with a beam convergence of 25 mrad. The inner and outer angles for HAADF-STEM imaging were 54-220 mrad.

### OER Measurements

The catalyst inks for the electrochemical measurements were prepared referring to methods reported by Suntivich *et al.*[12] and Grimaud *et al.*[13] $K^+$ ion-exchanged Nafion® was used as an immobilizing binder, enabling smooth transport of dissolved $O_2$ to the surface of the catalysts. A 3.33 wt.% $K^+$ ion-exchanged Nafion® suspension was first obtained from a mixture of 5 wt.% proton-type Nafion® suspension (Sigma-Aldrich) and a 0.1 M KOH aqueous solution in a 2:1 (*v/v*) ratio. This process increased the pH of the initial 5 wt.% proton-type Nafion® suspension from 1-2 to 11. The catalyst inks were prepared from a mixture of as-prepared samples (25 mg), acetylene black (AB, Strem Chemicals Inc., 5 mg) and 3.33 wt.% $K^+$ ion-exchanged Nafion® suspension (1.5 mL). The total ink volumes were adjusted to 5 mL by the addition of tetrahydrofuran (Sigma-Aldrich), to give final concentrations of 5 mg sample/mL, 1 mg AB/mL, and 1 mg Nafion/mL in the ink. A sample of the ink (6.4 μL) was then drop-casted on a rotating-ring disk electrode composed of a glassy carbon (GC) disk ($0.2 \times 0.2 \times \pi$ $cm^2$) and a Pt ring (BAS Inc., Japan), which was used as the working electrode after mirror polishing with 0.05 μM alumina slurry (BAS Inc., Japan). GC disk was dried overnight in vacuum at ~20 °C, and the catalyst layer was composed of 0.25 mg sample/$cm^2_{disk}$, 0.05 mg AB/$cm^2_{disk}$ and ~0.05 mg Nafion/$cm^2_{disk}$.

Electrochemical measurements were conducted using a rotating ring disk electrode rotator (RRDE-3A, BAS Inc., Japan) at 1600 rpm, in combination with ALS 701E electrochemical analyzer (BAS Inc., Japan). In addition, a Pt wire counter electrode and an Hg/HgO reference electrode (BAS Inc., Japan) filled with 0.1 M KOH (FUJIFILM Wako Pure Chemical Co.) were used. Electrochemical measurements were conducted with $O_2$ saturation (30 min $O_2$ gas bubbling) at ~20 °C, where the equilibrium potential of the $O_2/H_2O$ redox couple was fixed at 0.304 V vs. Hg/HgO (equivalent to 1.23 V vs. RHE). During the OER measurements for each sample, the potential of the sample-modified GC was controlled from 0.3 to 0.9 V vs. Hg/HgO (1.226-1.826 V vs. RHE) at 10 mVs$^{-1}$. During the ORR measurements, the potential of the sample-modified GC was controlled from −0.7 to 0.3 V vs. Hg/HgO (0.226-1.226 V vs. RHE) at 10 mVs$^{-1}$. For all measurements, the OER current densities were *iR*-corrected ($R$ = 43 Ω) using the measured solution resistance, and capacitance-corrected by averaging the anodic and cathodic scans to remove the non-faradaic current contribution.

### Density Functional Theory Calculations

DFT calculations were performed using the projector augmented-wave (PAW) method as implemented in the Vienna Ab initio Simulation Package (VASP)[56]. The exchange–correlation functional was treated using the Perdew–Burke–Ernzerhof (PBE) formulation within the generalized gradient approximation (GGA)[57] for both structural relaxations and electronic structure calculations. A plane-wave cutoff energy of 450 eV was used, and spin polarization was considered in all calculations.

For bulk calculations, the lattice parameters were fixed to experimentally reported values: $a = 10.4830$, $b = 5.8548$, $c = 8.2457$ Å for $NaAlB_{14}$, and $a = 10.3544$, $b = 5.8319$, $c = 8.1422$ Å for $AlB_{14}$. The Brillouin zone was sampled using a $3 \times 5 \times 4$ Monkhorst–Pack $k$-point mesh, and Gaussian smearing with $\sigma = 0.1$ eV was applied. Ionic positions were relaxed until the Hellmann–Feynman forces on all atoms were below 0.01 eV/Å, while keeping the lattice parameters fixed.

For surface calculations, slab models of $NaAlB_{14}$ and $AlB_{14}$ were constructed based on $1 \times 1 \times 2$ supercells with a vacuum layer of 15 Å along the c-axis. The Brillouin zone was sampled using a $3 \times 5 \times 1$ Monkhorst–Pack grid with the same Gaussian smearing. As shown in Figure S8, each slab contains four $B_{12}$ cluster units stacked along the c-axis. During structural relaxation, the two central units were fixed, while the two surface-facing units were fully relaxed.

### Acknowledgment
This work was supported by the Japan Science and Technology Agency (JST) CREST (Grant No. JPMJCR19J1), the Japan Society for the Promotion of Science (JSPS) (Grant Nos. 19H02420, 22H04458, 24K06950, 24K01593, 24K01171, and 25K08799). We would like to express our gratitude to Prof. Hiroshi Hirata, Graduate School of Information Science and Technology, Hokkaido University, for supporting ESR experiment, and Ms. Masae Sawamoto, Research Institute for Electronic Science, Hokkaido University, and Ms. Yukino Nishikubo, Multi-Materials Research Institute, National Institute of Advanced Industrial Science and Technology, for preparing the HPDC cells.

### Author Contributions
M.F. directed the research. M.F., H.M., and S.H. designed the experiments. H.M. synthesized the $NaAlB_{14}$ and $Na_2B_{29}$. M.F. performed the Na extraction from $NaAlB_{14}$ and $Na_2B_{29}$. S.H. measured the OER properties. M.J. conducted the TEM and STEM analyses. J.K.P. performed the XPS analysis. M.T. measured the magnetic and electrical properties. Y.I. performed the ESR analysis. M.F., K. M., and T. S. performed the DFT calculations. M.F. wrote the manuscript with input from all authors.

### Competing Interests

The authors declare no competing interests.

## Additional Information

Supplementary Information is available for this paper.

Correspondence and requests for materials should be addressed to Masaya Fujioka.

Supporting information

Boron Clusters for Metal-Free Water Splitting

Masaya Fujioka[1,2] *, Haruhiko Morito[3], Melbert Jeem[4], Jeevan Kumar Padarti[5], Kazuki Morita[6], Taizo Shibuya[7], Masashi Tanaka[8], Yoshihiko Ihara[9], Shigeto Hirai[5] **

[1] Multi-Materials Research Institute, National Institute of Advanced Industrial Science and Technology (AIST), Nagoya 463-8560, Japan
[2] Research Institute for Electronic Science, Hokkaido University, Sapporo 001-0020, Japan
[3] Institute for Materials Research, Tohoku University, Sendai 980–8577, Japan
[4] Faculty of Engineering, Hokkaido University, Sapporo, 060-8628, Japan
[5] Faculty of Engineering, Kitami Institute of Technology, Kitami 090–8507, Japan
[6] Helmholtz-Zentrum Berlin für Materialien und Energie, 12489 Berlin, Germany
[7] System Platform Research Laboratories, NEC Corporation, Kawasaki, Kanagawa 211-8666, Japan
[8] Graduate School of Engineering, Kyushu Institute of Technology, Kitakyushu 804-8550, Japan
[9] Department of Physics, Faculty of Science, Hokkaido University, Sapporo 060-0810, Japan

Corresponding authors
* m.fujioka@aist.go.jp
** hirai@mail.kitami-it.ac.jp

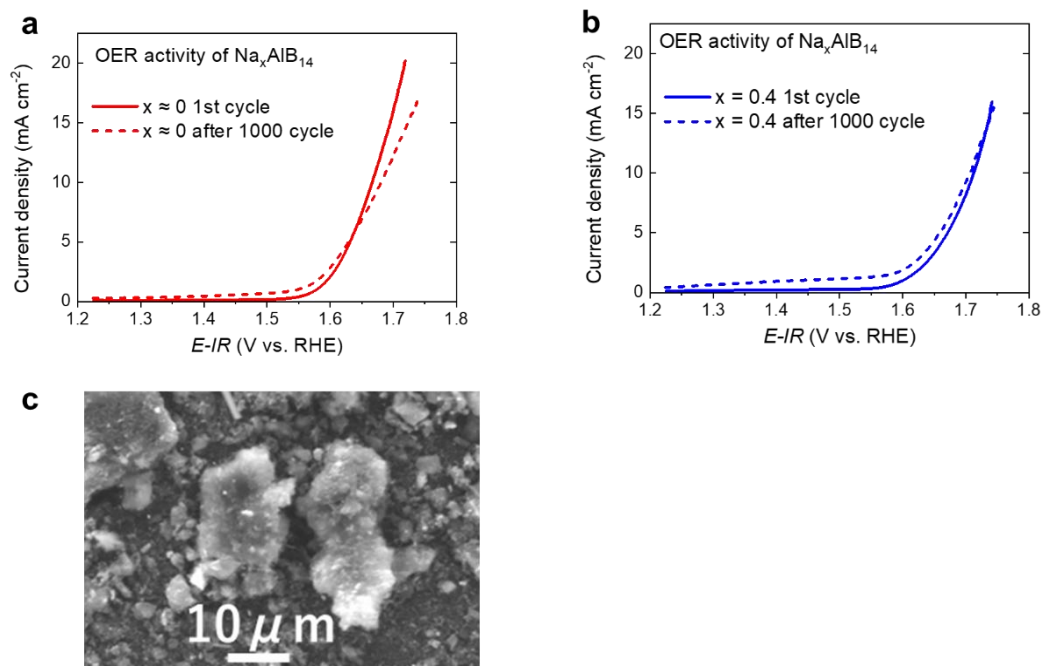

**Figure S1 | OER durability and morphology of Na$_x$AlB$_{14}$**

Initial and post-1000-cycle OER activity for Na$_x$AlB$_{14}$ with (a) x ≈ 0 and (b) x = 0.4. (c) SEM image of the sample after grinding.

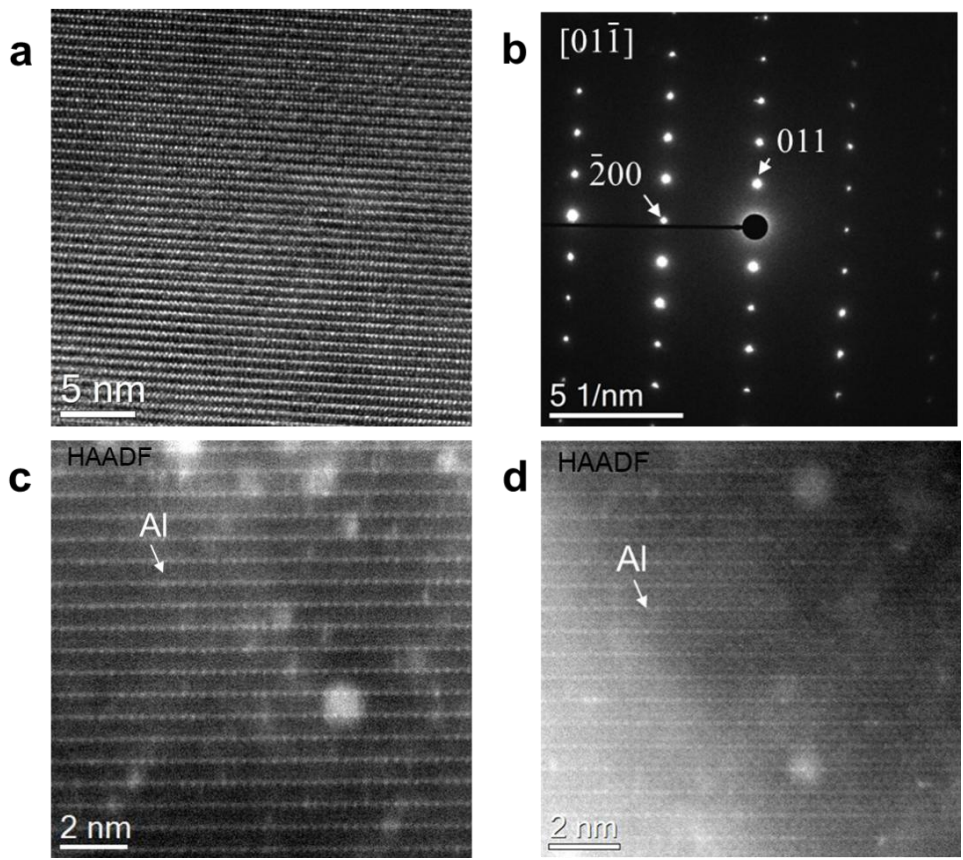

**Figure S2 | Structural integrity of Na$_x$AlB$_{14}$ before and after OER**

HRTEM and atomic resolution HAADF-STEM images of Na$_x$AlB$_{14}$ (x ≈ 0), focusing on the crystalline region. (a) HRTEM image before OER. (b) Selected area diffraction pattern of (a), obtained along [00$\bar{1}$] direction. Measured *d*-spacing were $d_{011}$ = 6.42 Å and $d_{\bar{2}00}$ = 2.92 Å. (c–d) Representative HAADF-STEM images before and after OER, respectively.

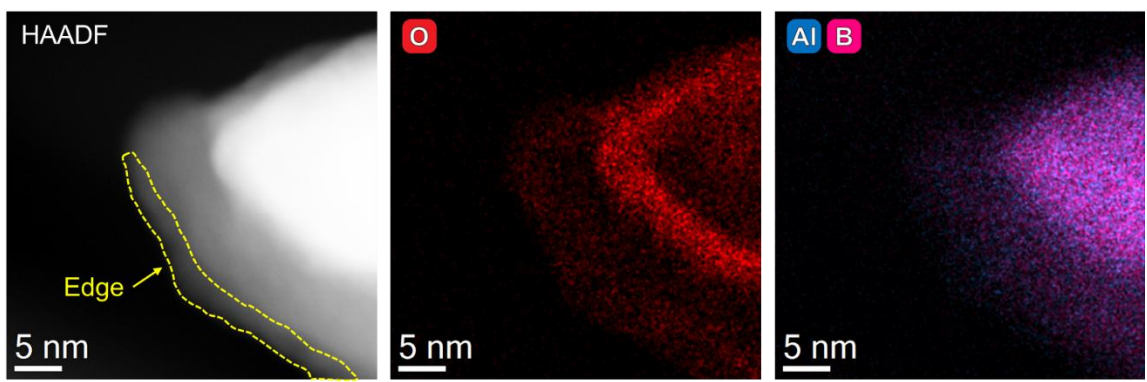

|  | Na Kα | Al Kα | B Kα | O Kα | total |
|---|---|---|---|---|---|
| Whole (at%) | 0 | 1.48 | 94.88 | 3.64 | 100 |
| Edge (at%) | 0.01 | 1.66 | 92.1 | 6.23 | 100 |

**Figure S3 | Elemental mapping of Na$_x$AlB$_{14}$ (x ≈ 0) after OER**

STEM-EDS maps and quantified atomic percentages of Na, Al, B, and O elements in Na$_x$AlB$_{14}$ (x ≈ 0) after 1000 cycles OER. The edge corresponds to the amorphous layer.

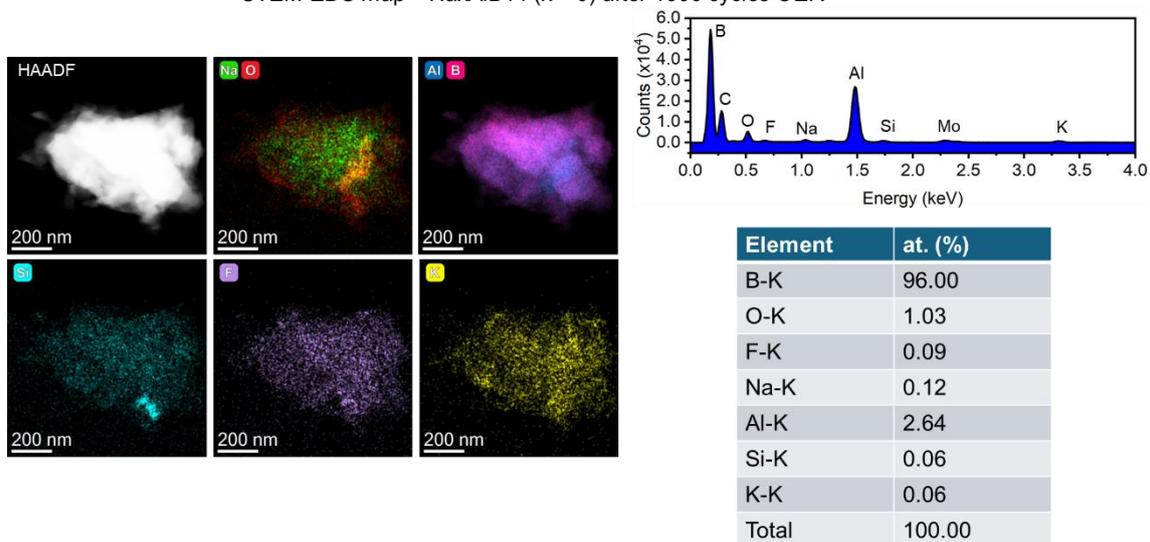

**Figure S4 | Additional EDS data for Na$_x$AlB$_{14}$ (x ≈ 0)**

STEM-EDS maps and quantified atomic percentages of Na$_x$AlB$_{14}$ (x ≈ 0) after 1000 cycles OER.

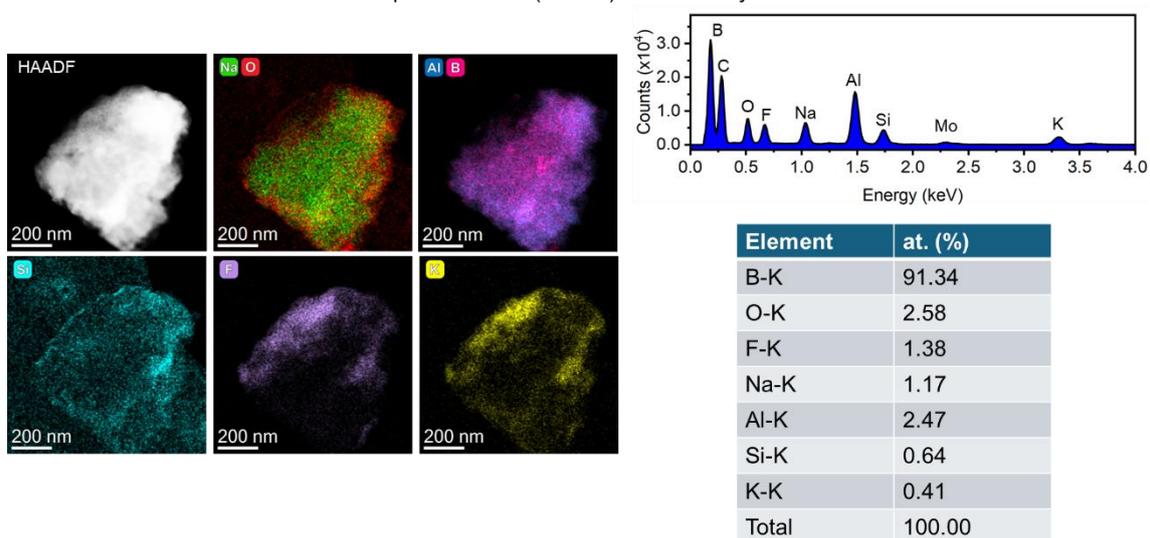

**Figure S5 | Additional EDS data for Na$_x$AlB$_{14}$ (x = 0.4)**

STEM-EDS maps and quantified atomic percentages of Na$_x$AlB$_{14}$ (x = 0.4) after 1000 cycles OER.

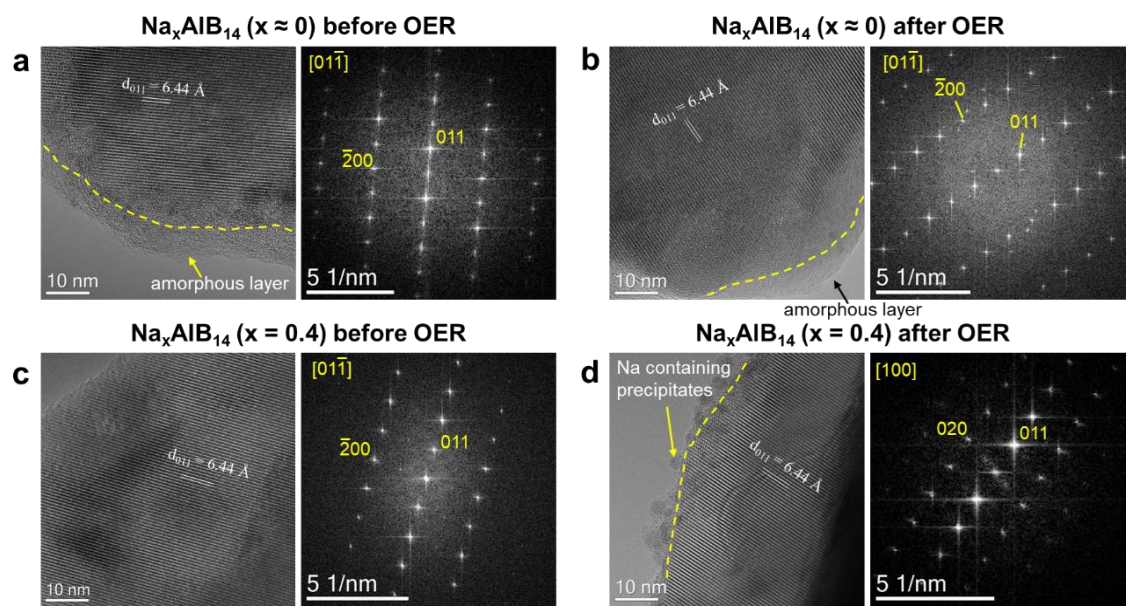

**Figure S6 | Surface structure evolution during OER**

HRTEM images of the Na$_x$AlB$_{14}$ samples and their corresponding FFT pattern. (a-b) x ≈ 0 sample before and after OER, respectively. (c-d) x = 0.4 before and after OER, respectively.

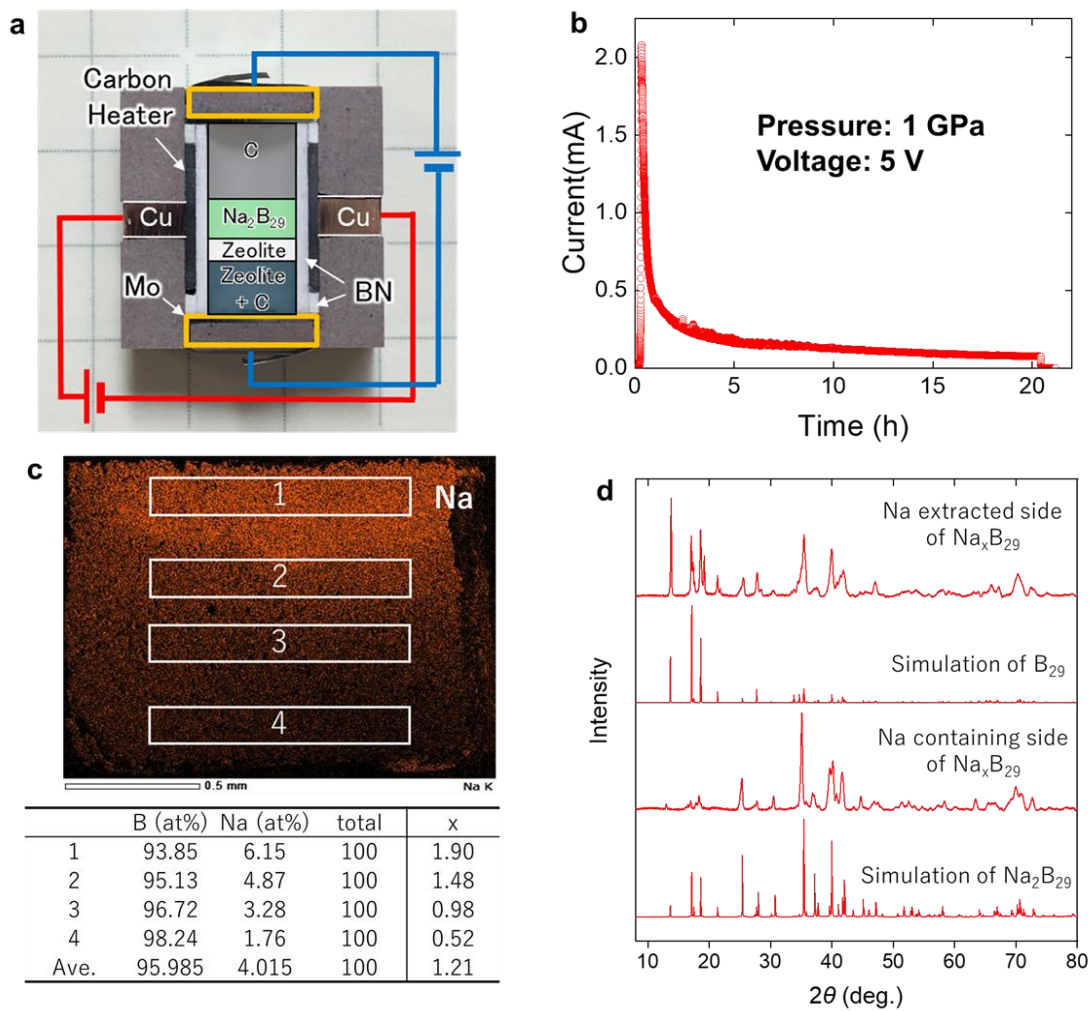

**Figure S7 | Na extraction from $Na_2B_{29}$ by HPDC**

(a) Schematic of the HPDC cell setup. (b) Time dependence of the current during Na extraction from $Na_2B_{29}$ under 1 GPa and 5 V. (c) STEM–EDS mapping image of the cross-section of $Na_xB_{29}$ and the corresponding atomic ratios. (d) XRD patterns of the zeolite side and carbon side of $Na_xB_{29}$ after HPDC treatment, compared with simulated patterns for $Na_2B_{29}$ and Na-extracted $B_{29}$.

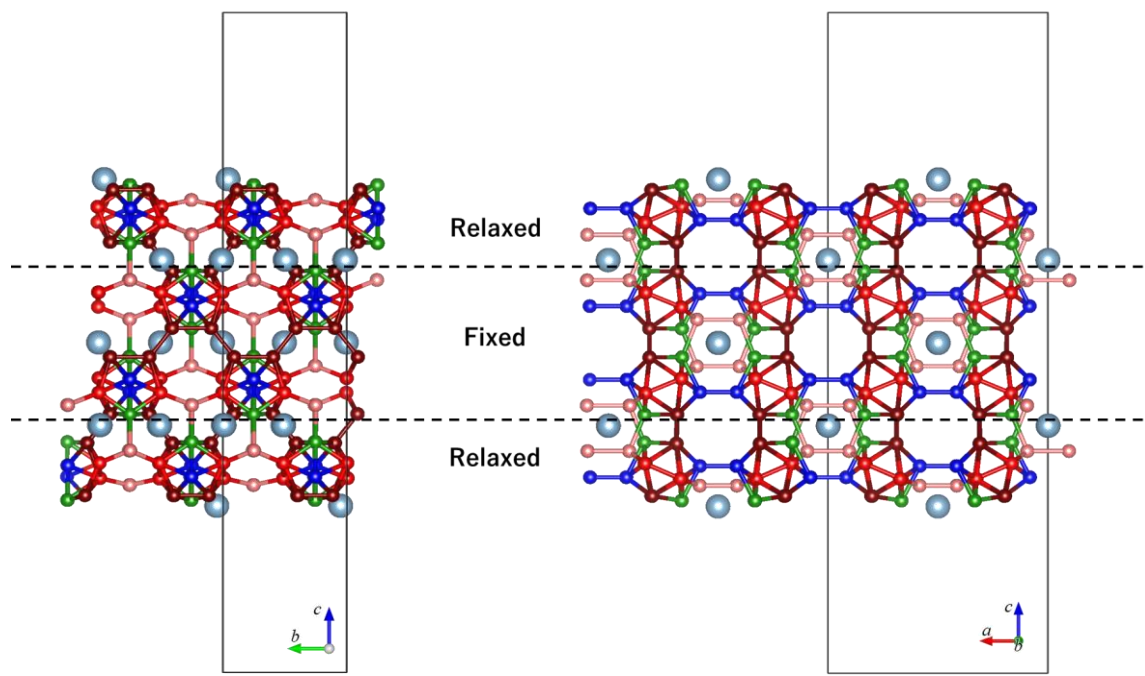

### Figure S8 | Slab model for DFT calculations

Slab model used for the DFT calculations. Structural relaxation was applied to the regions outside the dashed lines shown in the figure.